\documentclass{elsart}
\usepackage[dvips]{graphicx}

\begin{document}
\begin{frontmatter}
\title{Air fluorescence measurements in the spectral range 300-420 nm using a 28.5 GeV electron
beam.}

\author{R.~Abbasi,}
\author{T. Abu-Zayyad,}
\author{K. Belov,}
\author{J.~Belz,}
\author{Z.~Cao,}
\author{M.~Dalton,}
\author{Y. Fedorova,}
\author{P.~H\"{u}ntemeyer,}
\author{B. F Jones,}
\author{C.C.H.~Jui,}                              
\author{E.C.~Loh,}
\author{N. Manago,}
\author{K.~Martens,}
\author{J.N.~Matthews,}
\author{M.~Maestas,}
\author{J.~Smith,}
\author{P.~Sokolsky,}
\author{R.W.~Springer,}
\author{J.~Thomas,}
\author{S.~Thomas}
\address{Univ. of Utah, Salt Lake City, UT 84112, U.S.A.}

\author{P.~Chen,}
\author{C.~Field\corauthref{field}},
\corauth[field]{Corresponding author. Tel.: + 1-650-926-2694; fax: 
+1-650-926-4178}
\ead{sargon@slac.stanford.edu}
\author{C.~Hast,}
\author{R.Iverson,}
\author{J.S.T.~Ng,}
\author{A.~Odian,}
\author{K.~Reil,}
\author{D.~Walz}
\address{Stanford Linear Accelerator Center, Stanford University, 
Stanford, CA 94309, U.S.A.}

\author{D.R.~Bergman,}
\author{G.~Thomson,}
\author{A.~Zech}
\address{Rutgers Univ, Piscataway, NJ 08854, U.S.A.}

\author{F-Y.~Chang,}
\author{C-C.~Chen,}
\author{C-W.~Chen,}
\author{M.A.~Huang,}
\author{W-Y.P.~Hwang,}
\author{G-L.~Lin}
\address{CosPA, Taipei 106-17, Taiwan, R.O.C.}

\begin{abstract}
Measurements are reported of the yield and spectrum of fluorescence, excited by a 28.5 GeV
electron beam, in air at a range of pressures of interest to ultra-high energy cosmic ray detectors.
The wavelength range was 300 - 420 nm. System calibration has been performed using Rayleigh
scattering of a nitrogen laser beam. In atmospheric pressure dry air at 304 K the yield is $20.8
\pm 1.6$ photons per MeV.

\end{abstract}
 
\end{frontmatter}

\section{Introduction}

We report on a measurement of the fluorescent yield of air at wavelengths and pressures of
interest to large area cosmic ray shower detectors. Telescope arrays, imaging fluorescence in
large volumes of the atmosphere, continue to probe the spectrum of ultra-high energy cosmic
rays (UHECR). The present work is a step in establishing the foundation for the energy scale
calibration of UHECR observations.

The steeply falling spectrum is very difficult to measure above $\sim 10^{19}$ eV (1.6 Joules
per particle) because of the very low rates and our technical inability to produce and study
particles of such energies in the laboratory. It is of great interest \cite{Sigl}, however, because
conventional mechanisms do not explain the occurrence of such high energy particles, and this
holds out the possibility of uncovering new physics. One possibility is acceleration by very
energetic sources \cite{accel mech}. Alternatively the decay of primordial super-heavy particles
\cite{decay} might be responsible. Distinguishing between the mechanisms requires continued
statistical and systematic improvement in the measurement of the UHECR spectrum.

Recent experimental interest was focused on evidence for the Greisen-Zatsepin-Kuzmin (GZK)
\cite{GZK} cutoff in the spectrum, at about $10^{20}$ eV, expected from interactions with the
cosmic microwave background. The large ground level scintillation array AGASA \cite{AGASA
result} reported an unbroken spectrum above the limit. The atmospheric fluorescence array
HiRes \cite{HiRes result}, however, reported a steepening of the spectrum consistent with the
cutoff. This was recently confirmed by preliminary results from the nearly completed Auger
detector \cite{Auger-results}, whose energy scale is also set by fluorescence measurements. The
latter two arrays find that the spectral slope becomes shallower as energy increases above $\sim 5
\times 10^{18}$ eV, before a final steepening. But they do not yet agree on the flux, or the exact
energy thresholds and slope exponents of the segments of the spectrum. Uncertainty in the
knowledge of air fluorescence from cosmic ray showers may contribute to the differences. 
 
Building on previous experiments, increasingly precise measurements are anticipated for the
spectral shape and shower profiles above $\sim 10^{16}$ eV. This promises to afford a rich field
of study. Inferences about the relative contributions of galactic and extra-galactic fluxes, the
cosmic distance distribution of sources, and UHECR composition, will become accessible
\cite{C,L,M}. New experiments, at various stages of development, are aimed at enhanced
statistics with concentrated study of systematic effects \cite{Auger-TA-EUSO-OWL}. All of
these include at least a fluorescence measurement system for atmospheric showers, and depend
on an improved knowledge of the atmospheric fluorescence processes.

Fluorescence detectors can measure the cosmic ray energy by measuring the total light yield of
the shower that it generates in the atmosphere. From the initial interaction high in the
atmosphere, the secondaries interact again, still at high altitude, and as the air becomes denser, a
cascade quickly builds up, driven by the decay of the copious neutral pions produced. The decay
of the pions leads to the electromagnetic shower that, in fact, is the dominant vehicle for the
absorption in the atmosphere of the initial energy of the cosmic ray.

It is the electrons and positrons of the shower that transfer energy to the air molecules and cause
them to emit fluorescence light. The energy transfer is well described by the Bethe-Bloch
equation \cite{Bethe-Bloch}, and a large fraction comes from particles in the energy range of
tens of MeV. Within the gas molecules the energy flow is very complicated because of the
number of energy levels available, and the pressure-dependent competition between fluorescent
emission and energy transfer in molecular collisions. Work towards the theoretical modeling and
calculation of the fluorescent light yield is 
progressing \cite{Bunner,Blanco-Arqueros,Keilhauer,Arqueros-etal}.

The UHECR atmospheric fluorescence detectors use spherical mirrors to image the profile of the
light on to arrays of photomultiplier tubes \cite{HiRes result,Auger-TA-EUSO-OWL}. The
light's spectral range is restricted by filters to an ultraviolet window with low sky light
background between 300 and 400 nm. The fluorescence light in this wavelength range is
dominated by nitrogen emission lines, with major bands near 315, 337, 355, 380 and 391 nm,
(95\% of the light) and a few others of lesser intensity \cite{Bunner,Davidson_O'Neil}. It is
important to note that modern detectors can measure light from very high energy showers at
distances exceeding 30 km. But at such distances the Rayleigh scattering by air molecules
becomes important, and, even in the limited 300 to 400 nm range, its $\lambda^{-4}$
wavelength dependence preferentially reduces the detection efficiency of the short wavelengths.
At 337 nm, for example, the exponential scattering length at one atmosphere is 11 km. The
corrections for this (allowing for the vertical profile of atmospheric density) require a knowledge
of the spectrum of the emitted light.

The uncertainty on the fluorescence yield remains the largest single contribution to the overall
uncertainty. In order to match the advancing capabilities of the UHECR detectors, it is being
studied experimentally by several groups using different
techniques \cite{Nagano-2003,Nagano,Kakimoto,T-461,Colin,Airfly,Lefeuvre,FAFW}. The
yield and spectrum, as a function of atmospheric pressure, has been reported at several, often
quite low, electron energies. A study of the light yield as a function of depth in electromagnetic
showers, and the sensitivity of the spectrum to depth, has been reported \cite{Alumina}. The
present paper contributes light yield measurements over the range of pressures important for
UHECR showers, and a survey of the spectrum. It makes use of a detector calibration technique
systematically different from other approaches. The goal was to reduce systematic uncertainties
in the fluorescence yield and spectrum below 10\%, commensurate with other UHECR
experimental uncertainties.

\section{Experimental method}

Much of the work reported so far has used radioactive sources to excite the air in a test cell. This
corresponds to the low end of the dominant part of the shower's electron spectrum. The use of
pulsed high energy electrons entails a different set of systematic issues, and has some advantages.
For example, the monochromatic electron trajectories are easy to model, together with the
fiducial length for light emission from the test gas. With a pulsed beam, the light signals can be
strong, statistics may be collected relatively quickly, and photomultiplier tube random dark noise
does not contribute a background. On the other hand, linearity of the signal response must be
checked, heavy shielding is necessary from stray radiation, and backgrounds must be studied.

The beam available for this study was in the Final Focus Test Beam (FFTB) facility at the
Stanford Linear Accelerator Center. Electrons at 28.5 GeV were delivered at 10 Hz in pulses 3 ps
long. The apparatus was installed in an air gap in the beam vacuum line, with 50 micron thick
stainless steel vacuum windows upstream and downstream. The beam particle trajectories were
effectively parallel, and their transverse distribution was measured nearby using transition
radiation in visible wavelengths emitted from a titanium foil in the beam. This light was imaged
by a CCD camera and image capture system. Beam spot widths were typically $\sim$1 mm.

The value deduced for the fluorescence efficiency depends directly on the measurement of the
beam intensity. For this purpose, a toroid was mounted in the beam line upstream of the
fluorescence apparatus \cite{JNg-NIM}. The electron bunches passed nearly centrally through
this ferrite ring which had an evenly spaced copper winding.
The winding was coupled to front-end electronics, close to the beam line, which amplified the
current impulse and used a bandpass filter to improve noise rejection. This signal was sent to the
remote data collection system where it was digitized on every beam pulse. Beam intensities as
low as $10^7$ electrons per pulse could be measured with adequate resolution. At low intensity
the pulse-to-pulse intensity variation could be $\pm$30\%. The calibration of this unit, including
its linearity, was studied by injecting known charges into the electronics, and also, after removal
from the beam line, by inductive coupling into the toroid itself of fast pulses simulating beam
bunches. The calibration factor has been established with an overall uncertainty of better than
2.7\%. Since the toroid was upstream of the fluorescence apparatus, it did not sense the
perturbation on the electron flux caused by the thin beam windows. This has been evaluated by
simulation codes, as discussed below.

\begin{figure}
\begin{center}
\includegraphics*[width=7cm]{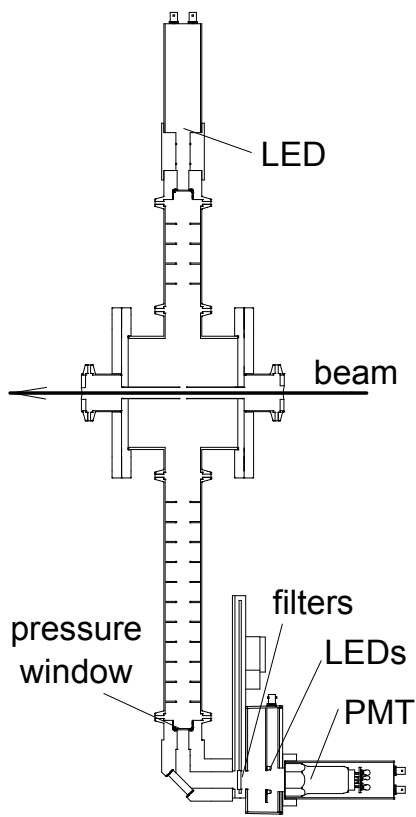}
\end{center}
\caption{Sectional view of the apparatus. The electron beam axis is indicated by the arrow. The
optical path is perpendicular to the electron beam along a baffled tube, through a fused silica
pressure window, to a right-angle reflection. Following this is a wheel of optical filters, LEDs
(outside the fluorescence light path) for monitoring stability, and then the PMT. The opposite
arm contains an on-axis LED.}
\label{apparatus}
\end{figure}

The apparatus in which the fluorescence occurred, and was measured, is illustrated in
Fig.~\ref{apparatus}. It consisted of a 25 cm long, 15 cm diameter cylindrical stainless steel
vessel, mounted coaxially with the electron beam. The upstream and downstream ends were
closed off by 25 micron thick aluminum pressure windows. Inside this, a pair of 1.6 cm diameter
thin aluminum cylinders were placed, one upstream, one downstream, coaxially with the beam,
with a gap of 1.67 cm between them in the center of the apparatus. This gap formed the defined
length of gas to be observed by the photomultiplier tubes. The internal surfaces were coated
black to suppress scattered light, including the forward emitted Cherenkov radiation from the
beam.

Light from the gap could pass down two light channels which extended out radially through the
cylinder walls. The channels were at right angles to each other, and were designated North and
South. Their interiors also were black, and had baffles to suppress unwanted light paths. Each
terminated, at 45 cm from the beam axis, in a 1.2 cm diameter fused silica pressure window that
formed the effective iris, as well as being the mechanical limit of the low pressure volume. The
light continued on for 15 cm, undergoing a right angle reflection at a UV enhanced aluminum
coated mirror. This turn in the optical axis allowed lead shielding to be placed between the
photon detector and the beam line, eliminating the relatively direct radial path for scattered
electrons or gamma rays to enter the face of the detector.

A wheel of optical filters was installed immediately after the right angle reflection. Various
narrow band filters were available, as was a sample of the 300 - 400 nm filter used for the HiRes
telescope, a clear aperture, and a blank position to study ``dark'' backgrounds. The rotation of the
filter wheels was controlled remotely.

The filtered light was collected by one photomultiplier tube, (38 mm diameter Photonis XP3062
\cite{Photonis}) in each light channel. The PMTs were samples from the inventory of tubes used
for the HiRes experiment. Their high voltages were monitored throughout the experiment.
Diametrically opposite each of these light channels was a shorter cylinder in which was mounted
an ultraviolet LED that was flashed between beam pulses to monitor PMT gain stability. Other
LEDs were placed (outside the fluorescence light path) between the filters and the PMT face.
Along with the PMTs, in the same shielding enclosure, was placed a similar tube with a hood
over its photocathode. This was intended to monitor noise not associated with fluorescent light,
particularly the effects of penetrating radiation. Although the apparatus, and in particular the
PMT section, was encased to the extent possible in lead shielding, some radiation could penetrate
and excite the PMTs, depending on beam conditions. An additional hooded tube with a different
gain was enclosed in a separate shielded cavity.

A gas system, with its controls outside the beam radiation enclosure, allowed the fluorescence
cylinder to be filled with dry air. For some measurements, ambient moist air was used to
investigate the effect of water vapor. For systematics studies, nitrogen or ethylene (which
fluoresces only very weakly in the relevant wavelength range) could be introduced. The system
pressure was varied in steps covering the range from 10 torr to 750 torr.

The PMT and toroid signals were recorded using a standard CAMAC gated
analog-to-digital-converter (ADC) system controlled by a PC. Also recorded were gas pressure,
filter position, PMT high voltages, and temperatures and gas humidity. During data collection,
occasional triggers were imposed to measure ADC pedestals and to pulse the LEDs used to
monitor the PMT gains.

\section{Optical calibration of the detector system}

Various factors contribute to the light detection efficiency: the geometrical optical acceptance;
the reflectivity and transmission of the optical materials; the PMT quantum efficiency; its
amplification (gain); and the calibration of cabling and the charge-to-digital conversion
efficiency of the ADCs. All the factors related to photons are wavelength dependent, although in
the range of interest, the optical reflectors and windows are almost negligibly so. In order to
measure all these factors in the intact optical set up, the system was moved to a laboratory where
the vacuum windows were replaced by coated optical windows. Where the electron beam had
been, the narrow beam from a nitrogen laser was aligned. The signal detected by the PMTs was,
in this case, the 337 nm laser photons that were Rayleigh-scattered from gas molecules in the
1.67 cm target gap. Since only $\sim{10^{-6}}$ of the photons scattered in the target gap at 1
atm., it was necessary to take great care to suppress stray light from elsewhere in the optical path.
The laser beam was monitored by a probe calibrated by the manufacturer \cite{power_meter} to
$\pm5\%$, and signals from the PMTs were measured with the system previously used at the
electron beam line. By comparing the PMT responses to the LED flashers in the laboratory with
those recorded in the electron beam enclosure, an average repeatability difference of 0.5\% was
measured, with a range from +2.5\% to -1.4\% for the different LEDs. For this contribution to the
calibration uncertainty, we conservatively take $\pm$2.5\%.

\begin{figure}
\begin{center}
\includegraphics*[width=14cm]{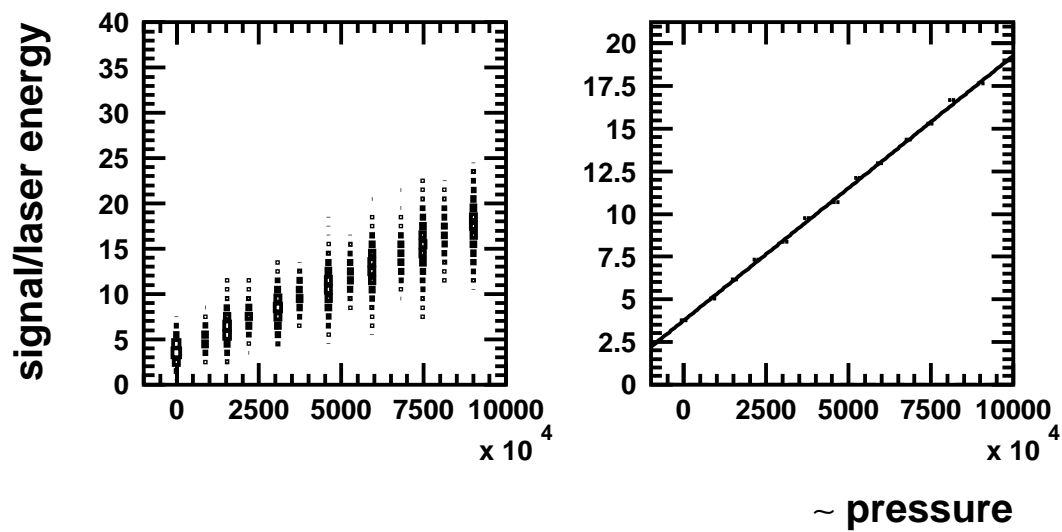}
\end{center}
\caption{PMT response against Rayleigh scattering intensity, controlled by changing air pressure.
The ordinates are ADC counts, corrected for pedestal and normalized by the laser power
measurement.}
\label{Rayleigh_v_press}
\end{figure}

Measurements were made at a wide range of gas pressures between atmospheric and vacuum. It
was found that the signal strength, normalized to the laser intensity, rose linearly with pressure,
Fig.~\ref{Rayleigh_v_press}, as expected from Rayleigh scattering. The intercept at the vacuum
setting corresponded to the background from errant laser rays, and the detected PMT signal from
Rayleigh scattering was obtained from the slope. The actual Rayleigh scattering yield was
calculated from the laser intensity and the gas density using the expressions provided by
Bucholtz \cite{Bucholtz}. A calculation based on Bodhaine et al. \cite{Bodhaine} gave results
consistent within 0.2\%.

Additional uncertainties taken into account include those of environmental changes on the gas
density (1.1\%), and of the straight line fit to the pressure curve (0.2\%). Measurements were also
made with the HiRes wide-band filter in place. The ratio of signals with : without filter gave a
value for its transmission at the 337 nm laser wavelength. This was also measured using a
spectrophotometer, and the discrepancy between the results of the two methods, $\pm 1.8\%$,
may be taken as an indication of the repeatability uncertainty.

\begin{figure}
\begin{center}
\includegraphics*[width=14cm]{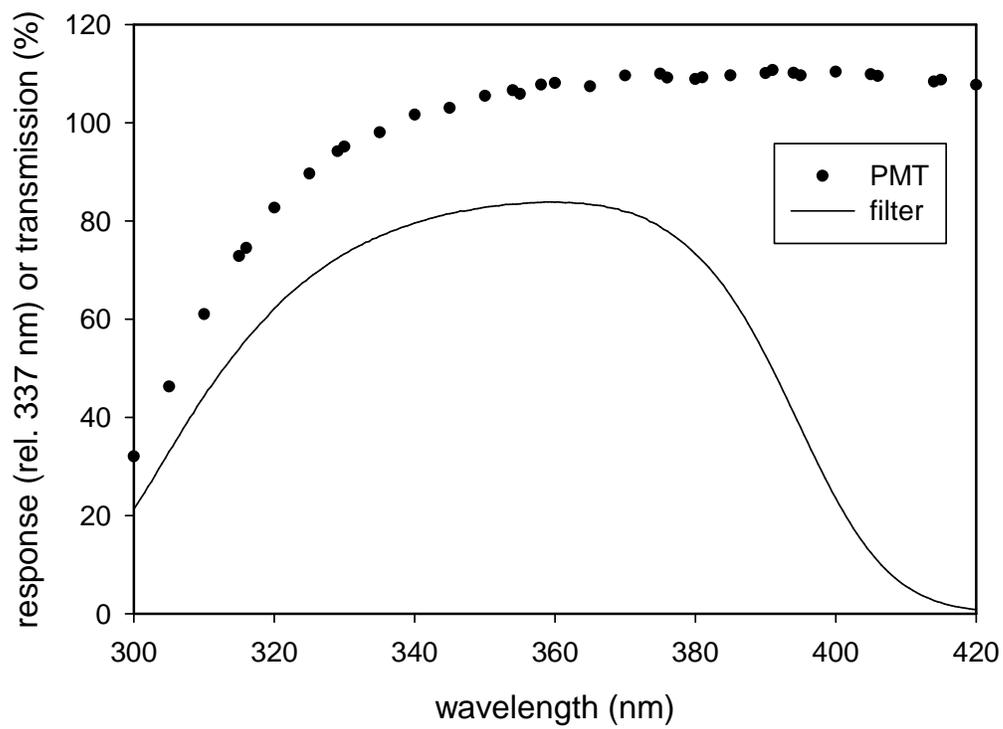}
\end{center}
\caption{PMT response relative to that at 337 nm, and transmission of HiRes filter, vs.
wavelength.}
\label{filter_line_shape}
\end{figure}

The calibration has been extended from 337 nm to the full wavelength range by comparing the
DC current responses of the PMTs and NIST wavelength-calibrated photodiodes \cite{PDs},
wavelength by wavelength, using the output from a monochromator, Fig.~\ref{filter_line_shape}.
Between 337 and 420 nm the PMT response was found to vary by less than 10\%. Below 337
nm, however, it dropped off, as expected for a bi-alkali photocathode. By 300 nm it was as low
as 30\% of the 337 nm performance. 
From wavelength to wavelength the rms fluctuations in the response measurement were always
less than 2\%.

In measurements without optical filters, a few percent more light may be detected beyond the
limits of the monochromator survey, and so the measured response function has been extended
using manufacturer's data. The overall sensitivity to the air fluorescence spectrum has been
computed using two examples of measured spectra. One was that reported by the Airfly
collaboration \cite{Airfly}, and the second, from this experiment, is discussed below. In both
cases, the spectrum was extended over the weak emission range to 600 nm, using the visible
wavelength results from Davidson and O'Neil \cite{Davidson_O'Neil}. The results of the
average response calculation for the two spectra agree within 0.1\% where a HiRes filter is used,
and within 3\% for the case with no filter. Relative to the 337 nm light, for air fluorescence the
system average response is 0.648 when the HiRes filter is used, and 0.995 without a filter, and
we assign an uncertainty of $\pm$0.015. To explore the effect of a possible shift in the
calibration of the monitor photodiodes across the width of the spectrum, a factor proportional to
the wavelength distance from the 337 nm calibration point has been introduced. By 420 nm the
imposed change in sensitivity was 5\%. Since most of the light is from wavelengths relatively
close to 337 nm, however, this changed the calculated response to the fluorescence spectra by
only 1.2\%, which is taken as a contribution to the overall uncertainty.
 
A correction was applied for the difference in geometry between the source of the light from
Rayleigh scattering and that from the more wide-spread energy deposition from the electron
beam. In the electron beam case, one source of loss is related to the limited transverse optical
acceptance of the light channels. Another effect is the suppression of low energy scattered
particles at larger radii, caused by the material of the beam tubes and the rest of the detector, an
effect that does not occur in atmospheric showers. The spread of the energy deposited by the
electron beam was simulated in a model of the apparatus using EGS4 \cite{EGS4}, and the
acceptance efficiencies at the optical iris were calculated numerically for both the laser and
electron beam cases. The factor obtained from Rayleigh scattering for the efficiency of
converting photons to ADC counts must be reduced by (3.2$\pm$0.25)\% for the case of the
electron beam in the apparatus fiducial volume. In addition, the ratio between the energy
deposited in a 1 cm length of free air without beam windows and that in the fiducial volume is
1.0837 $\pm$ 0.0015. For calculations using the actual energy deposit in the fiducial volume, an
overall simulation systematic uncertainty of 1\% should also be applied.

\section{Data analysis}

\subsection{Data processing and background subtraction}

\begin{figure}
\begin{center}
\includegraphics*[width=14cm]{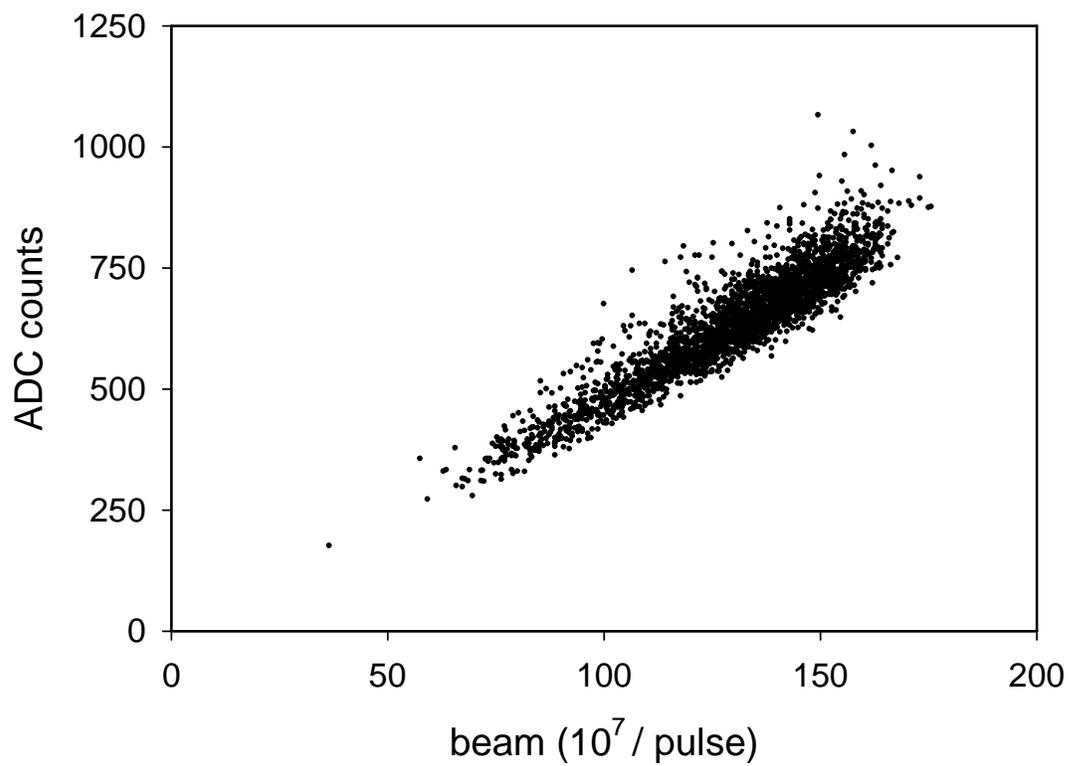}
\end{center}
\caption{Fluorescence signal plotted pulse-by-pulse against beam intensity at 750 torr, HiRes
filter}
\label{run-970}
\end{figure}

The data were accumulated in runs of several thousand beam pulses, and the gas pressure or
optical filters were changed between runs. Within each run, the ADC signals were corrected for
the zero-beam pedestals. The signals from the active PMT and the background counters were
plotted, pulse by pulse, against the measured beam charge, as in Fig.~\ref{run-970}. The slope of
the PMT versus beam regression is proportional to the fluorescent excitation caused by the
electron beam, together with a beam-related background contributed by radiation penetrating the
shielding and exciting the PMT.

For this work, only the results of the north PMT will be discussed. The south counter, with a
higher gain selected for use with narrow wavelength-band optical filters, was in saturation for the
configurations with relatively transparent filters used for overall sensitivity determination. The
narrow band filter results are intended to be the subject of a future report.

At the low beam intensities used, the delivered pulse charges typically were variable by $\sim\pm
30\%$. This permitted the signal vs. beam plot to be studied for systematic effects. For a
selection of runs with a suitable low beam intensity, where the PMTs were checked to be
responding linearly, the regression was found to intersect the beam axis at $(-0.41\pm 0.31)
\times 10^7$ electrons per pulse, consistent with zero. As a consequence, the origin has been
applied as a constraint on all subsequent fits, where the typical beam intensity was $\sim 10^9$.
To allow for the uncertainty, a conservative contribution of 1\% has been included in the overall
uncertainty evaluation.

\begin{figure}
\begin{center}
\includegraphics*[width=14cm]{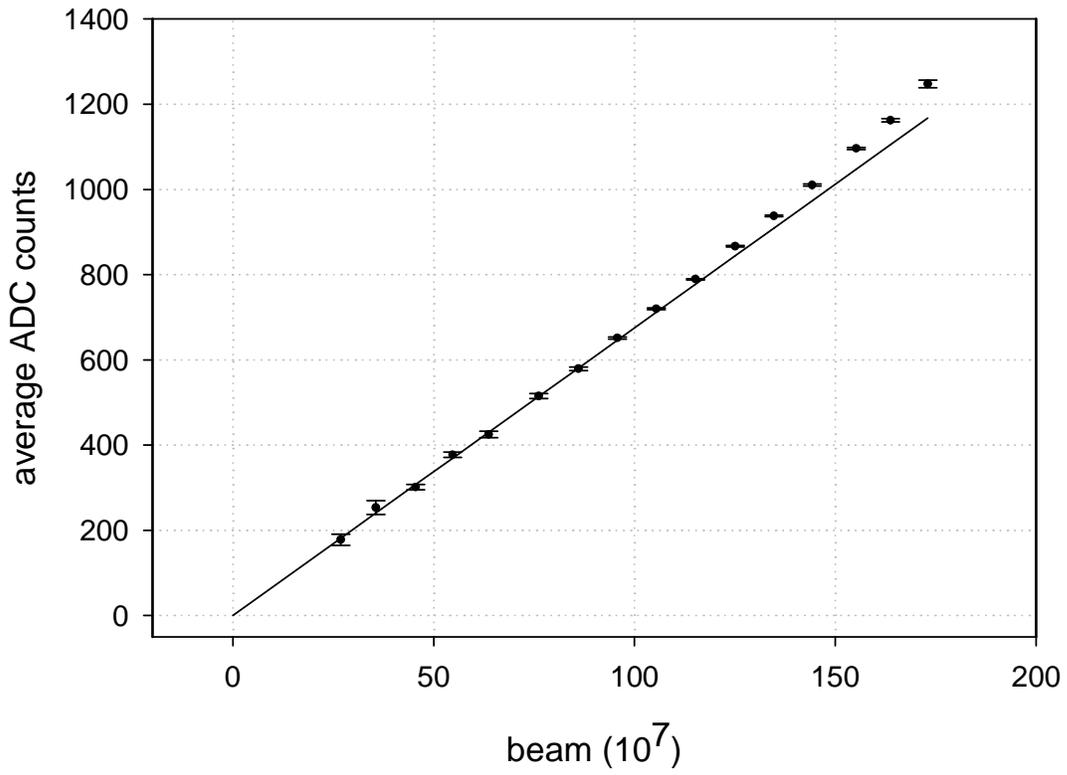}
\end{center}
\caption{Fluorescence plotted against beam intensity at 250 torr, no optical filter, showing the
enhancement occurring at higher beam intensities.}
\label{sig-enhancement}
\end{figure}

Examination of the data allowed the limits of linearity to be determined. Two effects have been
observed. The normal ``saturation'' sensitivity fall off of the PMT was seen to affect a small
fraction of high pressure data where no optical filter was used. This was avoided by imposing a
cut on beam intensity in runs showing this effect. A second, and more important, effect has its
origin in non-linear effects caused by the collective electric field impulse of the intense electron
bunch. This field has the ability to accelerate electrons from ionization events, and these in turn
cause molecular excitations, adding  to the fluorescence signal. Related effects have been studied
in the same beam line \cite{E-150}. This enhancement of the fluorescence
is illustrated in Fig.~\ref{sig-enhancement}. We find that the beam pulse intensity threshold for
the enhancement decreases with gas pressure, from $1.5\times 10^9$ at 1 atm to $0.8\times
10^9$ at 50 torr, for beam spots $\sigma_x \times \sigma_y$ of 1 mm $\times$ 1 mm, and 1 mm
long. Beam intensity cuts were made to avoid this problem, and, as a result, some higher intensity
runs could not be used.

The backgrounds caused by radiation penetrating the shielding were measured directly. This was
achieved by rotating the filter wheel to the opaque position and fitting against beam intensity as
before. Runs in this condition were interspersed among the others. Short term variations in the
background were accounted for by monitoring the signals of the two permanently hooded
``background'' PMTs. Although their pulse-to-pulse fluctuations were large, over run-length
intervals their averages tracked each other well. They have been averaged to provide a run-to-run
adjustment of the opaque-filter background runs. These backgrounds, typically $\sim 5\%$ for
data without a filter, or $\sim 7\%$ with the HiRes filter, were then subtracted from the results of
the neighboring fluorescence run fits.

There were electron beam runs in both 2003 and 2004, and the Rayleigh scattering calibration
was performed after the latter. The stability of the system between the two beam runs was tested
by comparing sets of data taken under similar conditions, and the variations observed lead us to
assign an uncertainty of $\pm 2\%$ for the long term stability.

\subsection{Photon yield in dry air}

With any given setting of pressure and filter, several runs were taken, separated in time by hours
to days. At each setting, the results show very good consistency, and so they have been averaged.
The variation between them has been used to estimate a run-to-run uncertainty, which is included
in the final results.

After applying the calibration discussed above in the optical calibration section, the dry air
results are given in Table~\ref{phot/MeV} for some pressures in the range of interest for
UHECR detectors. Only HiRes filter data are available at 50 torr.

The uncertainties that should be applied to these figures are listed in Table~\ref{uncer} in units
of percent. They have been discussed in the narrative above. 

From these, the overall uncertainty on the yield per MeV is 7.5\%: i.e. at 760 torr the yield is 20.8
$\pm$ 1.6 photons per MeV.

\begin{figure}
\begin{center}
\includegraphics*[width=14cm]{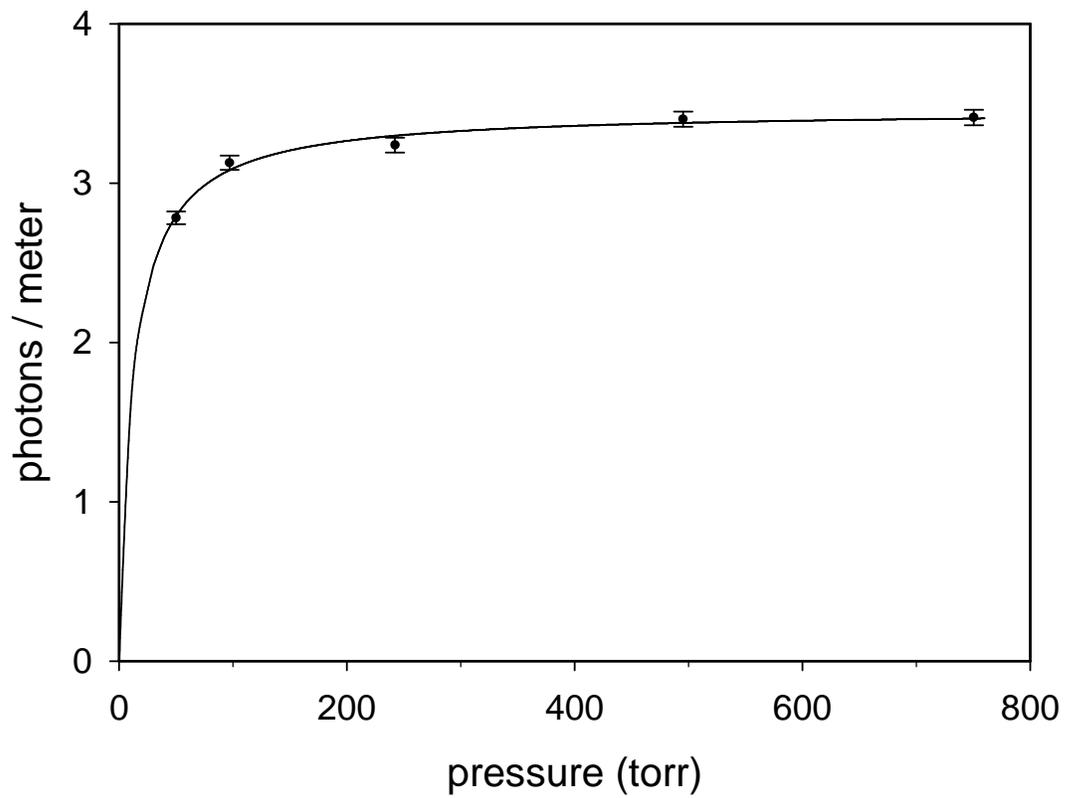}
\end{center}
\caption{Photons per meter measured using the HiRes filter. The error bars shown indicate the
point-to-point uncertainty of 1.42\%. An overall scale uncertainty of 7.5\% applies to all points.
See the text for the fit expression.}
\label{ph_per_m}
\end{figure}

Results have frequently been quoted in terms of photons per meter, the electron energy loss per
meter being a function of its energy. We also give these values in Table~\ref{phot/m} for both no
filter and HiRes filter cases. The HiRes filter results are also shown in Fig.~\ref{ph_per_m}. For
air, the fluorescent yield of an electron track barely increases with increasing pressure above
$\sim 0.1$ atm. This is because of increasing competition from molecular collision processes,
principally involving oxygen, which compensate for the increased energy deposit with gas
density. The fit line shown is the expression 
$a P / (1+b P)$ \cite{Bunner,Kakimoto} motivated by quenching of the fluorescent molecules by
pressure dependent collisions. Here $P$ is pressure and $a, b$ are constants. The error bars
shown in this case are the point-to-point relative uncertainties of 1.42\%. The overall 7.5\% scale
uncertainty still applies.

The data were taken at a temperature of 304 K. The effect of temperature $T$ is to change the
collision rate with the quenching oxygen molecules, represented as a modification in the
denominator of the above expression:  $a P / (1+b P (T / 304)^{1/2})$
\cite{Bunner,Kakimoto}. The increase in the signal would, for example, be 1.3\% at the 296 K
temperature reported in Ref \cite{Colin}. At 273 K it would be 5.4\% , and would reach 10\%
only at 250 K.

\subsection{Nitrogen, ethylene and humid air}

As a systematic check, some data were collected with nitrogen in place of air. Collisional
de-excitation is much weaker in pure nitrogen than in air. In the absence of heavy optical
filtering, the stronger nitrogen signal at higher pressures caused non-linearity in the PMT at all
but the weakest beam intensities. In response, we have fitted a quadratic function to the nitrogen
data. The fits were repeated for various beam intensity ranges up to limits between 1.2 and 1.5
$\times 10^9$, and the variation in the linear term was noted. This variation was taken as an
indication of the systematic error of this approach, 4\% for data without an optical filter, 0.2\%
with the HiRes filter. Additional uncertainties came from evidence of a small manifold leak of
dry air into the nitrogen (1.7\%) and from the signal attenuator (4\%). It has been found that, at
750 torr, the signal ratio between nitrogen and dry air is $6.51 \pm 0.37$ with no optical filters,
or $6.84 \pm 0.29$ with the HiRes filter. This is in agreement with our previously reported
\cite{T-461} value of $6.6 \pm 0.2$. 

Results from a short run with a fill of ethylene were processed in the standard way, but the
signals were barely above background. At 750 torr, the strengths relative to air were $(0.34 \pm
0.13) \%$ with the HiRes filter, and somewhat larger, $(0.88 \pm 0.09) \%$ without a filter. This
can be used to set an observational limit on the Cherenkov light background.

Ethylene yields 2.5 times as much Cherenkov light as air. If all the ethylene signal were from this
process, then its contribution to the air signal, at 750 torr and with the HiRes filter, would be
$(0.138 \pm 0.052) \%$, or $< 0.21 \%$ at $90 \%$ confidence. The wider wavelength range of
the no-filter case would not increase the Cherenkov signal by more than $\times 1.4$ relative to
the HiRes filter case. This is already negligible for our purposes, but it is likely that much of the
weak ethylene signal is the result of fluorescence. From its design parameters, the black beam
tube is expected to suppress the Cherenkov light contribution at 1 atm, the worst case, to $<
0.1\%$ of the air signal.
                              
Filtered air from outside the building, containing a relatively large fraction of water molecules,
was also studied. At the overall pressure of 750 torr, the partial vapor pressure of water was 11
torr, or 1.5\%. The data taking and processing were similar to those for dry air data. The results
show that, at 750 torr the humidity reduced the light yield by $(7.4\pm 0.3)\%$, consistent
between HiRes- and no-filter cases. Data from a single no-filter run and a HiRes filter run, with
room air at 245 torr, were also available. In these cases the yield reductions were $(3.5 \pm
1.4)\%$ and $(8.8 \pm 2.6)\%$. We do not consider the discrepancies to be significant, and so
take a weighted average of both pressures. That is, 1.5\% water vapor content (equivalent to
7.1\% of the oxygen partial pressure) suppressed the signal by $(7.3 \pm 0.3)\%$. By
comparison, at the temperature of the standard atmosphere at 400 torr or 5000 meters altitude,
the water vapor partial pressure saturates at 0.25\%.

\section{Spectrographic observations}

During part of the data taking, an opportunity arose to use an independent arrangement in the
beam line to record emission spectra with relatively high resolution. This was accomplished by
mounting an 8 cm long, thin-windowed, gas cell in the beam, and observing the fluorescence
light emitted perpendicularly to the beam through an aperture 2.6 cm along and 1.3 cm transverse
to the beam. The gas cell was baffled to suppress scattered Cherenkov light. 
Two aluminum-coated $45^{\circ}$ plane mirrors, a fused silica pressure window, and a
focusing mirror were used to match the light from the cell to the f/4 acceptance of the 120 mm
focal length spectrograph \cite{spectrograph} in a heavily shielded enclosure. Signals were
measured pulse by pulse by using a multi-anode photomultiplier tube with a linear array of 32
pixels spaced at 1 mm \cite{hamamatsu}. The anode signals were digitized by the standard
CAMAC 11-bit ADCs. Beam intensity information was not available synchronously for this data
set, and so the data does not support comparisons of total yield at different times, although the
shapes of spectra may be compared.

Between runs the light path was deflected away from the spectrograph slit, and this gave a
background measurement for off-line subtraction. The raw signals were averaged over each run
taken at a fixed pressure, and the background, including the ADC pedestal, was subtracted. The
plot of these values already, at this raw stage, showed the characteristic spectrum of the gases
studied, dry air (including argon and carbon dioxide) and nitrogen. Unpopulated gaps between
the expected spectral lines excluded the presence of a significant continuum spectrum such as a
Cherenkov radiation background.

After the beam run, the wavelength calibration of the system was made by measuring the
positions of the lines from a mercury discharge lamp, and the sensitivity as a function of
wavelength was studied by recording its performance when illuminated by the standard
continuous spectrum of a deuterium lamp. It was found that the response of the system as a
function of wavelength was dependent on the light path through the spectrograph. When using a
calibrated light source, it was not possible to match the distribution in space and angle of the
beam-induced fluorescence light, and this has led to an uncertainty of $\sim 15\%$ in the relative
light yield of the short wavelengths $\sim$ 315 nm, relative to the rest of the spectrum which
extended to 415 nm.

The analysis corrected for crosstalk between neighboring anodes, given as 3\% by the
manufacturer, consistent with our observations. Additionally, there was evidence for a loss of
linearity of anode signals for the strong lines, also consistent with the manufacturer's
specifications. A correction for this was evaluated using pulse amplitude data for the strong lines
at 337 nm and 357 nm and the line at 391 nm, using different pressures and a wide range of
approximately known average beam intensities. Most of the significant lines in our spectral range
are from neutral nitrogen excitations of the second positive (2P) system $C^3\Pi_u \rightarrow
B^3\Pi_g$. The non-linearity correction for the 2P lines was found to rise with pressure to a
plateau above $\sim$ 300 torr. This was expected, since the light flash width has been observed
to become narrower inversely with pressure \cite{T-461}, thereby increasing the peak dynode
current. The 391.4 nm band, labeled 1N(0,0), is a transition between vibrational levels
$B^2\Sigma^+$ and $X^2\Sigma^+$ of ionized nitrogen. For the 1N line, the pulse width had
been observed to have already become narrow by 50 torr, and, as this led us to expect, the non-
linearity correction rose more quickly with pressure to a constant level. For stronger lines at
higher pressures, the non-linearity corrections were $\sim 10\%$. An average systematic
uncertainty of 17\% of the correction was assigned.

\begin{figure}
\begin{center}
\includegraphics*[width=14cm]{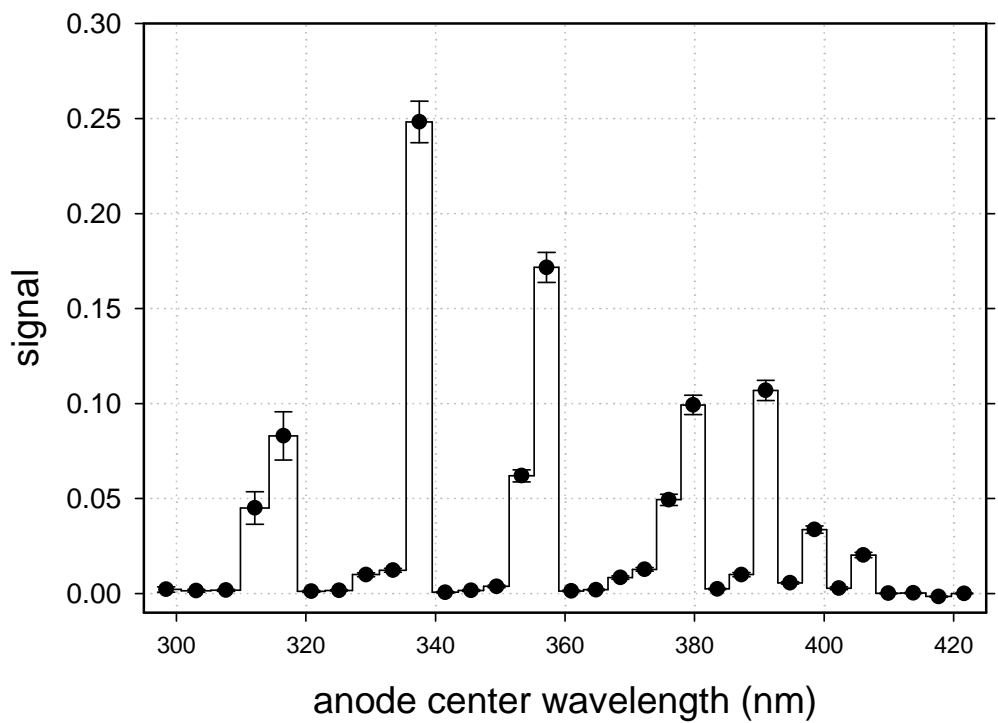}
\end{center}
\caption{Fluorescence spectrum of 155 torr air as seen by the multi-anode PMT. Corrections for
transmission, conversion efficiency and non-linearity have been applied. The sum of entries is
normalized to unity.}
\label{spc1}
\end{figure}

\begin{figure}
\begin{center}
\includegraphics*[width=14cm]{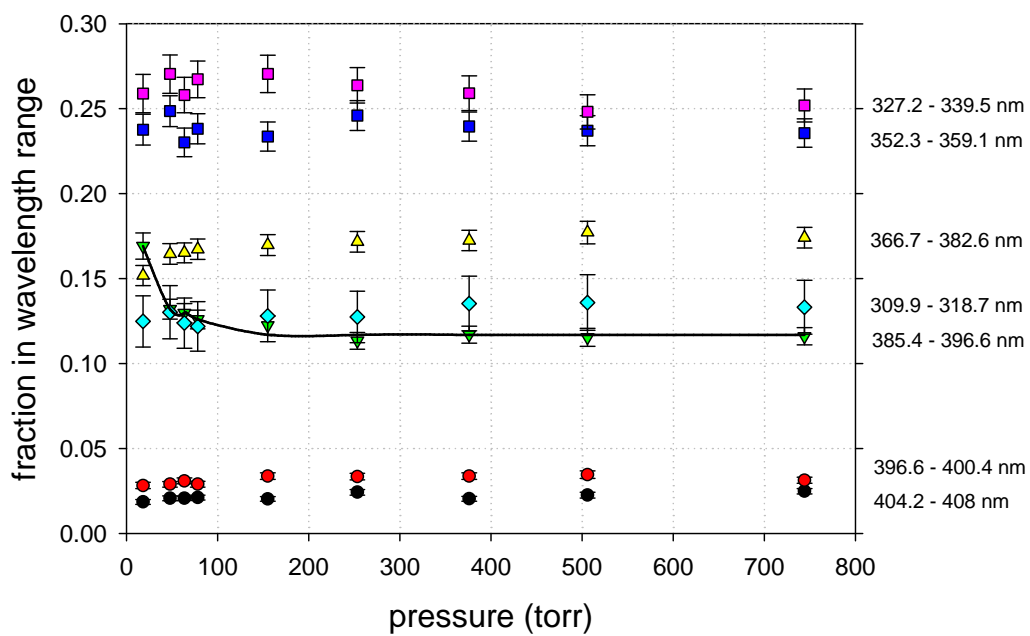}
\end{center}
\caption{Pressure variation of the relative contributions of various spectral ranges. The most
noticeable structure is the rise below 60 torr of the fraction of light in the 391 nm 1N band,
illustrated by the line.}
\label{spc2}
\end{figure}

An example of a spectrum, from air at 155 torr, corrected in this way is given in Fig.~\ref{spc1}.
Each bin corresponds to one anode and the sum of the bins is unity. In Fig.~\ref{spc2} are
shown, for air at various pressures, the relative signal strengths within several wavelength ranges
incorporating the most important lines. At each pressure, the corrected signals are expressed as a
fraction of the total signal at that pressure. There is not a strong change in the spectrum over the
pressure range of interest (above $\sim$100 torr). The transmission efficiencies of these spectra 
through a HiRes filter vary by 0.46\% rms, the largest deviation from the mean being 0.70\%.
Also, the Rayleigh scattering transmission efficiencies through 30 km of 1 atm. clear air varies by
less than 1\% for the spectra at pressures above 60 torr. Associated with its increased decay time,
the 391 nm transition of the 1N system is seen to become relatively more important as pressure
decreases below 60 torr. This is expected theoretically \cite{Keilhauer} because of differences in
the radiative and collisional de-excitation rates of this quantum state. 

In order to translate from a ``spectrum" of anode signals to one of emission lines, an additional
step was required. This was because the sensitivity profile across the 1 mm width of each anode
rose and fell between the inter-anode gaps. It followed an elliptical shape, averaging a 20\% loss
from uniform sensitivity, while dispersed images of the slit could fall into poor acceptance
regions, increasing uncertainties. A parameterization of the manufacturer's average acceptance
profile for this was included in a numerical model of the device. This permitted the alignment
and slit width calibrations to be applied and varied within the estimated uncertainties.

\begin{figure}
\begin{center}
\includegraphics*[width=14cm]{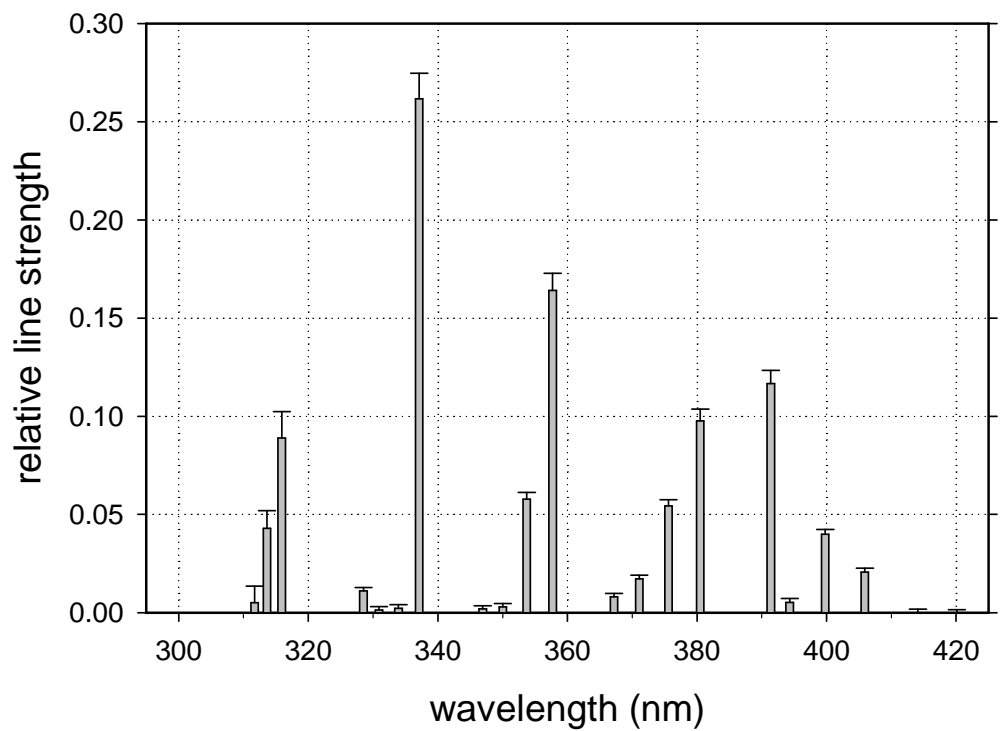}
\end{center}
\caption{Line spectrum from 155 torr air, obtained by matching known lines with the response of
the spectrograph. The sum of the line strengths is set to unity.}
\label{spc3}
\end{figure}

The air fluorescence spectra for resolved emission lines between 300 and 415 nm were extracted
from this simulation. The result at 155 torr, approaching the pressure of the upper altitude range
of interest to UHECR detection, is illustrated in Fig.~\ref{spc3}, where the signals are expressed
as a fraction of the total yield in the full wavelength range. At this pressure, 
the photomultiplier tube non-linearity corrections, and their uncertainties, are relatively small.

It may be of significance that the optical system focused a region of just $\pm 2$ mm around the
electron beam on to the spectrograph slit, and so excluded fluorescence from energy deposited
beyond this. It appears possible that the spectra from the core region of the beam and the
peripheries may be somewhat different \cite{Arqueros-etal}. The 391.4 nm band, labeled
1N(0,0), is caused dominantly by direct ionization by the high energy beam, while the 2P system,
responsible for the other important lines, is excited by low energy scattered electrons. With the
tightly selected field of view, the 391 nm region may have been detected more efficiently than
the rest of the range.

In data runs taken with a lower resolution grating, a comparison of dry air and room air, with
1.5\% partial pressure of water, showed no discernable humidity related change in the spectrum
profile.

\section{Discussion}

Several points are worthy of attention. The sensitivity calibration of the photomultiplier system
with the HiRes filter has an uncertainty better by a factor of $\sim 2$ than has been available
before. The change in the HiRes UHECR spectrum caused by this is within previous
uncertainties, as will be reported elsewhere.

\begin{figure}
\begin{center}
\includegraphics*[width=14cm]{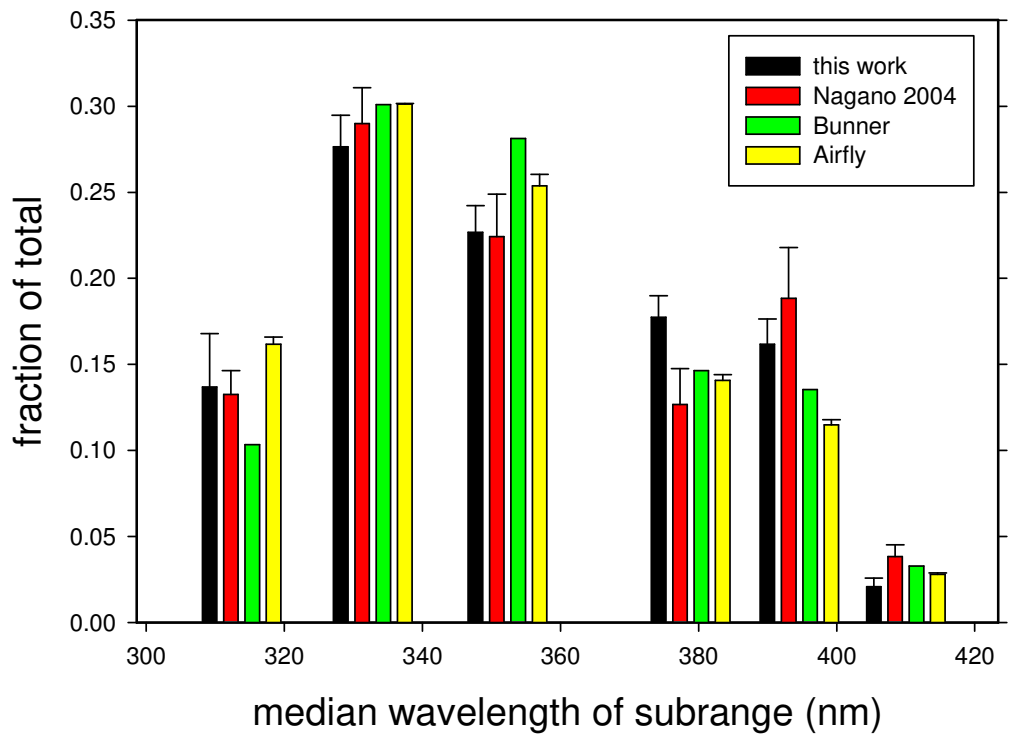}
\end{center}
\caption{Comparison between published emission strengths and those reported here, grouped
into six wavelength bands. Minor adjustments have been made to allow for compatible
wavelength ranges. In each case the sum of emissions is normalized to unity.}
\label{D1}
\end{figure}

It is also reassuring that the observed spectral structure of the emitted light indeed comes from
the expected nitrogen bands \cite{Bunner,Davidson_O'Neil}. In comparing with spectrum
measurements reported elsewhere, one finds general agreement. There are discrepancies in detail,
however, reflecting the difficulty and range of parameters and techniques involved.
In Fig.~\ref{D1} are illustrated various reported signals in wavelength bands suitable for the
UHECR detector range 300 to 420 nm. Reports from the early work of Bunner \cite{Bunner}, the
optical filter work of Nagano and collaborators \cite{Nagano}, the Airfly collaboration
\cite{Airfly} and from this paper, are shown for comparison in the same bands. The values
plotted have been expressed as a fraction of the total signal reported within the range.

The agreement is adequate for present UHECR data analysis. The transmission efficiency of the
HiRes optical filter for these spectra is the same within 1\%. We have compared the effect on
them of the wavelength dependence of Rayleigh scattering. The transmission values for these
four spectra, calculated through a 10 km air column at 1 atm., average 0.481, with a worst case
deviation from the mean of 0.009. At 20 (30) km atm the average transmission is 0.242 (0.124)
with a worst case deviation of 0.008 (0.009). (The average pressure along a light ray from the
peak of a shower to the UHECR detector is typically 0.6 - 0.7 atm.) At 30 km atm, these
differences, amounting to up to 7\% of the transmitted signal, will become of significance for
future detectors, and will then need clarification. 

\section{Acknowledgements}

     We thank the staff of the accelerator operations group for their skill in delivering the
unusual beam so efficiently, and the personnel of the CEF Dept. for their
dedication and hard work in preparing and installing the infrastructure of this experiment. We
also thank the technical staffs at the universities for their substantial efforts. The work was
supported in part by the U.S. Department of Energy, contract DE-AC03-76SF00515 and by the
National Science Foundation under awards NSF PHY-0245428, NSF PHY-0305516, NSF PHY-
0307098 and NSF PHY-0400053.


\clearpage
\begin{table}
\begin{center} 
\caption{Photons per MeV as measured with no filter and penetrating the HiRes-filter. An overall
uncertainty of 7.5\% applies to the yields (see text)}
\begin{tabular}{c c c}
\hline
& \multicolumn{2}{c}{photons / MeV}\\
      \cline{2-3}
pressure (torr) & no filter & through HiRes filter \\
\hline
\hline
 760 & 20.8 & 14.0 \\
 495 & 32.0 & 21.7 \\
 242 & 64.3 & 43.0 \\
 97 & 157.6 & 105.2 \\
50 & & 182.2 \\
\hline
\end{tabular}
\label{phot/MeV}
\end{center}
\end{table}

\clearpage
\begin{table}
\begin{center}
\caption{Contributions to uncertainty on the photon yield, in units of percent}
\begin{tabular}{ l c }
\hline
uncertainty contribution & \% \\
\hline
 beam calib. & 2.7 \\
signal splitter & 1 \\
zero constraint of fits & 1 \\
run-to-run stability & 1 \\
laser vs e-beam light source shape & 0.4 \\
simulation & 1 \\
spectrum sensitivity, open filter & 1.5 \\
spectrum sensitivity, HiRes filter & 1 \\
beam line vs lab stability & 2.5 \\
2003 data calib. & 2 \\
filter consistency checks & 1.8 \\
PMT relative spectral response & 1.2 \\
Rayleigh scattering: & \\
laser power & 5 \\
gas density for laser scattering & 1.1 \\
theoretical calculations & 0.2 \\
fit slope & 0.2 \\
\hline
\end{tabular}
\label{uncer}
\end{center}
\end{table}

\clearpage
\begin{table}
\begin{center}
\caption{Photons per meter for no filter and HiRes filter cases. The relative uncertainty between
points at different pressures is 1.42\%, and an overall scale uncertainty of 7.5\% applies to all
points.}
\begin{tabular}{ c c c }
\hline
pressure (torr) & no filter & HiRes filter \\
\hline
\hline
750 & 5.059 & 3.413 \\
495 & 5.029 & 3.403 \\                                      
242 & 4.848 & 3.240 \\
97 & 4.686 & 3.128 \\
50 & & 2.784 \\
\hline
\end{tabular}
\label{phot/m}
\end{center}
\end{table}

\clearpage

\clearpage

\clearpage

\clearpage

\clearpage

\clearpage

\clearpage

\clearpage

\clearpage

\clearpage

\end{document}